# Performance of Localized Coupled Cluster Methods in a Moderately Strong Correlation Regime: Hückel−Möbius Interconversions in Expanded Porphyrins

Nitai Sylvetsky,[†] Ambar Banerjee,[†] Mercedes Alonso,* and Jan M. L. Martin*

**ABSTRACT:** Localized orbital coupled cluster theory has recently emerged as a nonempirical alternative to DFT for large systems. Intuitively, one might expect such methods to perform less well for highly delocalized systems. In the present work, we apply both canonical CCSD(T) approximations and a variety of localized approximations to a set of flexible expanded porphyrins— macrocycles that can switch between Hückel, figure-eight, and Möbius topologies under external stimuli. Both minima and isomerization transition states are considered. We find that Möbius(-like) structures have much stronger static correlation character than the remaining structures, and that this causes significant errors in DLPNO-CCSD(T) and even DLPNO-CCSD($T_1$) approaches, unless TightPNO cutoffs are employed. If sub-kcal mol$^{-1}$ accuracy with respect to canonical relative energies is required even for Möbius-type systems (or other systems plagued by strong static correlation), then Nagy and Kallay's LNO-CCSD(T) method with "tight" settings is the suitable localized approach. We propose the present POLYPYR21 data set as a benchmark for localized orbital methods or, more broadly, for the ability of lower-level methods to handle energetics with strongly varying degrees of static correlation.

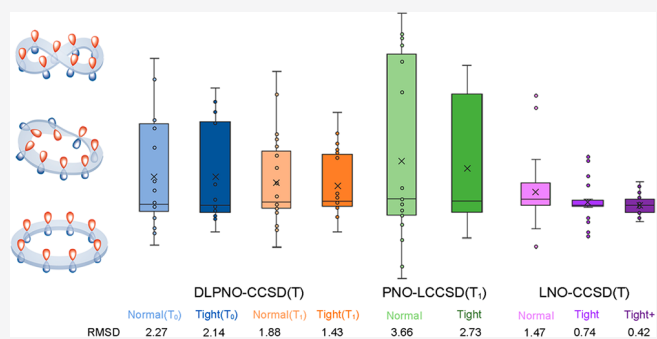

## ■ INTRODUCTION

Expanded porphyrins have drawn much attention over the past few decades due to their facile redox interconversions, novel metal coordination behaviors, versatile electronic states, and conformational flexibility.[1] The latter is responsible for the rich chemistry associated with such systems, which has led to various applications such as near-infrared dyes,[2] nonlinear optical materials,[3] magnetic resonance imaging contrast agents,[4] and molecular switches.[5] Contrary to the parent porphyrin, expanded porphyrins are flexible enough to easily undergo conformational changes, which correspond to distinct π-conjugation topologies (Hückel, Möbius, and twisted-Hückel/figure-eight) encoding different chemical and physical properties (Scheme 1).[6,7]

Such changes may involve a Hückel−Möbius aromaticity switch within a single molecule, which can easily be induced by, inter alia, an appropriate solvent, pH, temperature, and metalation conditions.[8,9] Thus, these Hückel−Möbius aromaticity switches have already been recognized for their potential applications in molecular optoelectronic devices.[10] Additional applications for expanded porphyrins, including conductance switching devices[11,12] and efficient nonlinear optical switches,[13] have recently been covered in the literature.

In a very recent collaboration[6] with Alonso et al., relative energies and isomerization pathways of a set of expanded porphyrins were investigated using wave function *ab initio* methods and DFT methods,[6] motivated by the fact that DFT-based energetics were shown to be highly dependent on the density functional employed in the calculations.[14,15] Furthermore, different DFT studies on expanded porphyrins have arrived at contradictory findings concerning the best-performing functionals to be used for these systems.[14−18] Indeed, since the stability of these isomers depends on the complex interplay of different factors (hydrogen bonding, π···π stacking, steric effects, ring strain, aromaticity, and so forth), it is no surprise that the selection of an exchange-correlation functional appropriate for describing the energy profiles of such topological switches is no trivial task. Thus, in ref 6, we opted to assess the performance of different exchange-correlation functionals for describing the thermochemistry and kinetics of topology interconversions in $N$-fused [24]-penta-, [28]hexa-, and [32]heptaphyrins—by comparing them to benchmark results obtained at the canonical CCSD(T)/CBS level of theory. The structures included in this benchmark are illustrated in Figure 1.

**Scheme 1. Representation of Different π-Conjugation Topologies of Expanded Porphyrins and Their Expected Aromaticities as a Function of the Number of π-Electrons**

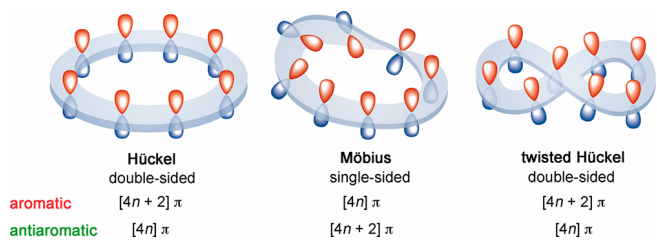

|  | Hückel double-sided | Möbius single-sided | twisted Hückel double-sided |
|---|---|---|---|
| aromatic | [4n + 2] π | [4n] π | [4n + 2] π |
| antiaromatic | [4n] π | [4n + 2] π | [4n] π |

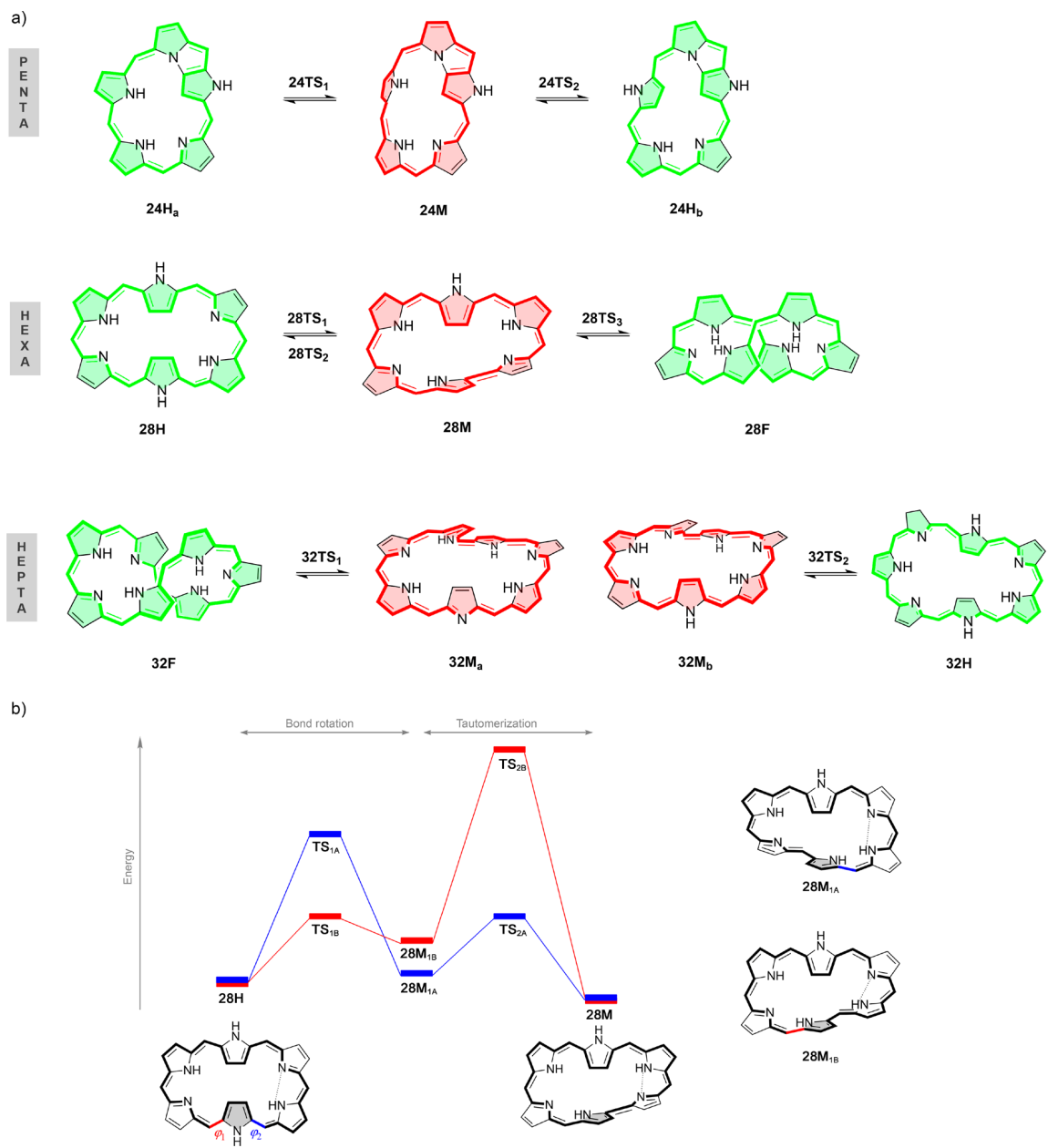

**Figure 1.** (a) Hückel (**H**), Möbius (**M**), and figure-eight (**F**) conformations of selected expanded porphyrins. Aromatic and antiaromatic macrocycles are colored in red and green, respectively. (b) Two **28H** ⇌ **28M** interconversion pathways investigated for the Hückel−Möbius interconversion in [28]hexaphyrin.

Unfortunately, canonical CCSD(T) calculations are notorious for their heavy computational burden, having formal CPU-time scaling of $O(n^3N^4)$, $n$ being the number of electrons in the system and $N$ corresponding to the number of basis functions employed in the calculation. Hence, even for heptaphyrins with the cc-pVDZ basis set, canonical CCSD(T) hit the ceiling of our computational resources. As an illustration, a canonical CCSD(T)/cc-pVDZ calculation on structure **28M** required no less than two node months total CPU time. Thus, treating even larger polypyrroles by means of robust, nonempirical *ab initio* methods is only feasible using alternative, computationally more economical methodologies.

DLPNO-type approaches (domain localized pair natural orbital), which have recently gained popularity due to their near-linear scaling properties, embrace the notion of pair natural orbitals (PNOs) in order to reduce the virtual space which has to be taken into account in a given calculation.[19−22] Recent methodological developments have led to the situation in which, using modern commodity servers, systems with over 44,000 basis functions and 2,300 atoms[23] are within reach of localized orbital-based *ab initio* methods. They may therefore constitute an obvious solution for the larger expanded porphyrins.[24,25]

That being said, the systems under consideration are known to be strongly delocalized:[26,27] thence, one may intuitively expect that localized orbital-based correlation approaches, such as the above-mentioned DLPNO-type ones, would prove to be inadequate. For this reason, assessing the performance of DLPNO-type approaches against canonical benchmark results is essential for confirming their reliability for highly delocalized $\pi$-systems.

We shall therefore assess the performance of several localized orbital approaches for the problem at hand. Below we show that some of the structures, specifically Möbius $\pi$-systems and the transition states resembling them, suffer from elevated degrees of static correlation, and that the errors for such systems can reach several kcal mol$^{-1}$ for the more cost-effective localized methods, although such errors can be mitigated through judicious choice of cutoffs.

## ■ THEORETICAL METHODS

In the present work, we shall consider four different localized orbital approaches. The first and second, both used as implemented in ORCA 4.1 and later, are two variants of the MPI-Mülheim DLPNO approach. The popular DLPNO-CCSD(T) approach, in which off-diagonal Fock matrix elements are neglected in the (T) contribution (such elements vanish for closed-shell canonical orbital calculations, but not for localized orbitals), actually corresponds to an approximation to canonical CCSD(T$_0$).[28] The latter approximation is eliminated in the more rigorous DLPNO-CCSD(T$_1$)[29] approach, at considerable additional CPU cost and I/O overhead.

The third method is the PNO-LCCSD(T) approach of Werner and co-workers[30,31] as implemented in MOLPRO 2018.[32] It likewise eschews the (T$_0$) approximation but differs substantially from DLPNO-CCSD(T) in terms of domain construction strategy—as explained in detail in refs 23 and 24 and summarized below.

Finally, we consider the LNO-CCSD(T) approach of Kállay and co-workers[23] as implemented in the MRCC package.[33] Here, the correlation energy is partitioned into occupied orbital contributions, and domains are adjusted for each such orbital individually to ensure that it is adequately represented. For orbitals that are not strongly delocalized, domains will be small, while strongly delocalized orbitals will entail large domains. As we shall see, this mitigates errors in such cases. For example, in the present work and for the systems at hand, we found that Möbius structures of the hexaphyrin required LNO-CCSD(T) wall times a factor of 4−5 longer than for simpler Hückel structures, compared to only about a factor of 2−2.5 for DLPNO-CCSD(T).

Each of the above DLPNO, PNO, and LNO methods has an array of cutoffs, screening thresholds, and other numerical parameters too unwieldy for routine manipulation by the nonspecialist user. Hence, typically several tuned combinations of such settings are offered that aim to consistently yield a given numerical precision for optimal computational cost. In the case of DLPNO-CCSD(T) in ORCA,[34] for example, three ascending levels of accuracy are collected under the keywords LoosePNO, NormalPNO (the default), and TightPNO: for details see Table 1 of ref 34. NormalPNO aims to yield energetics precise to 1 kcal mol$^{-1}$, while TightPNO sets the bar higher and is intended for applications like noncovalent interactions or conformer/isomer equilibria, where 1 kcal mol$^{-1}$ would be an unacceptably large fraction of the interaction and relative conformer/isomer energies, respectively. Similarly, PNO-LCCSD(T) in MOLPRO offers "Normal" and "Tight" domain settings (cf. Tables 1−4 of ref 31), while the corresponding MRCC settings are detailed in Table 1 of Nagy and Kállay.[35]

While the DLPNO-CCSD approach in ORCA and the equivalent PNO-LCCSD method in MOLPRO are very similar in their fundamentals and both achieve roughly linear CPU time scaling with system size, they differ considerably in their practical implementation details. Aside from the subtle differences in screening and cutoff strategies between codes, one more fundamental variance has chemical consequences for highly delocalized systems. Both codes construct virtual orbital domains for each correlation pair from the PAOs (projected atomic orbitals, i.e., the original basis set after projecting out all occupied MO components) and then constructs virtual orbital "domains" from these for the diagonal pair correlation $E_{ii}$ of each localized MO $i$ [domains for off-diagonal pair $E_{ij}$ are taken as the union of the domains for the diagonal pairs $E_{ii}$ and $E_{jj}$]. Pair natural orbitals are then calculated at the MP2 level, and only those PNOs whose natural orbital occupation number exceeds a set threshold are retained.

*Where DLPNO-CCSD(T1) in ORCA, and PNO-LCCSD(T) in MOLPRO, differ is how domains are constructed.* MOLPRO uses a spatial criterion based on a fixed number of atom shells (or a given maximum distance) around the bonded atom pair, viz., the atom that the lone pair sits on.[31,36] In contrast, ORCA uses an orbital population (older version) or orbital overlap (newer version) based criterion. In the older version,[20,22] all atoms for which the orbital had a Mulliken population greater in absolute value than the population cutoff parameter TCutMKN were included in the domain; whereas in ORCA 4 and later,[37] the orbital is included if the square root of the differential overlap is greater than the differential overlap cutoff parameter TCutDO. The MOLPRO approach typically yields much more compact domains, while the ORCA approach appears to be more resilient toward highly delocalized systems, such as the considered polypyrroles. It should be noted that, for nonconjugated molecules, the two approaches may be expected to perform comparably well.

Ma and Werner[31] have argued that, in view of the much faster basis set convergence of F12 approaches, their ultimate goal is PNO-LCCSD(T)-F12: the deficiencies of the smaller PNO domains would then in practice be obviated by inclusion of F12 corrections. While we acknowledge this argument, we do not currently have a viable way of generating canonical CCSD(T)-F12 data for such large systems. Canonical CCSD(T) reference data for a DZP basis set, on the other hand, proved computationally tractable albeit demanding. We do believe that it is valuable to test the approximations in the localized methods in isolation against the corresponding canonical answers, our view "uncluttered" by any F12 correction.

How do specific domain size settings affect the CPU time required for a given calculation? As an example, we consider the **28M$_{1B}$** Möbius structure. For DLPNO-CCSD(T$_1$)/cc-pVDZ, the CPU time ratio TightPNO:NormalPNO was 8.65:1; in other words, the more lenient settings save ∼88% of the total CPU time required for such a calculation. A somewhat smaller ratio (5.81:1) was observed for the DLPNO-CCSD(T$_0$) calculation. All else being equal, we find that DLPNO-CCSD(T$_1$) for these systems requires about twice the CPU time for DLPNO-CCSD(T$_0$): the fairly high I/O bandwidth required for (T$_1$) required running on nodes with high-speed local SSD (solid state drive) scratch disk volumes. For the problem at hand, it may be said that neither approaches require outlandish computational resources—and that the difference between them is still small enough to justify "going the extra mile" for superior accuracy.

Table 1. Post-CCSD(T) Corrections (kcal mol$^{-1}$) for the Relative Energies of [24] N-Fused Pentaphyrin, [28]Hexaphyrin, and [32]Heptaphyrin structures[a]

| system | CCSD(T) | ICE-CI | CCSD(T) | ICE-CI | CCSD(T) | ICE-CI | CCSD(T) | ICE-CI | CCSD(T) |
|---|---|---|---|---|---|---|---|---|---|
| active space | all orbitals | (12,12) | (12,12) | (18,18) | (18,18) | (24,24) | (24,24) | (30,30) | (30,30) |
| 24H$_a$ | 9.12 | 6.79 | 6.82 | −0.53 | −0.49 | 4.84 | 4.84 | 4.49 | 4.36 |
| 24H$_b$ | 0.00 | 0.00 | 0.00 | 0.00 | 0.00 | 0.00 | 0.00 | 0.00 | 0.00 |
| 24M | 6.06 | 8.12 | 8.24 | 4.89 | 4.96 | 7.90 | 7.92 | 8.40 | 8.31 |
| 24TS$_1$ | 9.05 | 6.70 | 6.71 | 3.29 | 3.28 | 6.68 | 6.62 | 6.53 | 6.38 |
| 24TS$_2$ | 4.87 | 6.00 | 6.04 | 3.08 | 3.09 | 5.86 | 5.83 | 6.39 | 6.27 |
| 28H | 0.00 | 0.00 | 0.00 | 0.00 | 0.00 | 0.00 | 0.00 | 0.00 | 0.00 |
| 28M | −0.73 | 10.88 | 11.12 | 8.91 | 9.09 | 9.28 | 9.43 | 7.62 | 7.56 |
| 28M$_{1A}$ | 0.46 | 12.60 | 12.87 | 9.91 | 10.12 | 10.39 | 10.54 | 8.55 | 8.48 |
| 28M$_{1B}$ | 1.82 | 13.57 | 13.82 | 11.67 | 11.86 | 10.98 | 11.09 | 11.38 | 11.34 |
| 28F | −0.38 | 7.41 | 7.38 | 9.18 | 9.12 | 5.40 | 5.28 | 4.70 | 4.45 |
| 28TS$_{1A}$ | 6.33 | 13.75 | 13.77 | 12.24 | 12.16 | 10.82 | 10.66 | 14.06 | 13.92 |
| 28TS$_{1B}$ | 2.86 | 9.14 | 9.12 | 10.02 | 9.97 | 8.72 | 8.60 | 6.56 | 6.16 |
| 28TS$_{2A}$ | 6.87 | 26.41 | 26.68 | 28.09 | 28.31 | 24.60 | 24.74 | 22.21 | 22.05 |
| 28TS$_{2B}$ | 9.89 | 30.33 | 30.57 | 31.42 | 31.62 | 28.31 | 28.44 | 26.44 | 26.30 |
| 28TS$_3$ | 5.17 | 15.03 | 15.02 | 14.55 | 14.44 | 13.31 | 13.15 | 12.17 | 11.84 |
| 32F | 0.00 | 0.00 | 0.00 | 0.00 | 0.00 | 0.00 | 0.00 | 0.00 | 0.00 |
| 32M$_a$ | 16.81 | 18.35 | 18.63 | 10.75 | 10.99 | 13.47 | 13.71 | 12.55 | 12.62 |
| 32M$_b$ | 16.74 | 18.46 | 18.72 | 13.85 | 14.15 | 15.71 | 16.03 | 17.48 | 17.98 |
| 32H | 34.60 | 22.91 | 22.90 | 24.75 | 24.74 | 24.18 | 24.17 | 27.23 | 27.49 |
| 32TS$_1$ | 17.49 | 16.64 | 16.64 | 10.69 | 10.67 | 11.23 | 11.22 | 14.13 | 14.39 |
| 32TS$_2$ | 33.79 | 24.28 | 24.22 | 24.71 | 24.65 | 25.40 | 25.37 | 27.58 | 27.74 |
| RMSD[a] | - | | 0.15 | | 0.14 | | 0.13 | | 0.21 |
| MUE[a] | - | | 0.10 | | 0.10 | | 0.09 | | 0.16 |

[a] See Figure 1 for the structural notation. RMSD and MUE (in kcal mol$^{-1}$) for the relative energies computed with ICE-CI and CCSD(T) methods for different orbital active spaces. Orbitals in the (n,n) active space are the n/2 highest occupied molecular orbitals and the n/2 lowest unoccupied MOs, selected from HF level orbital energies.

The CPU requirements stand in stark contrast to those for the corresponding *canonical* calculations, which are almost 2 orders of magnitude larger: as said above—running massively parallel on eight 16-core machines with a fully nonblocking InfiniBand interconnect and local SSD (solid state disk) scratch on all machines, canonical CCSD(T) on **28M$_{1B}$** required about 1 week total wall clock time. Moreover, adding just one more pyrrole ring into the macrocycle already quadruples the required time for the canonical calculation, while the difference is barely noticeable in the DLPNO or PNO calculations. Formally, canonical CCSD(T) asymptotically scales with system size $n$ as $O(n^7)$, while DLPNO-CCSD(T) and PNO-LCCSD(T) asymptotically scale linearly.

As part of the present work, we have also considered the following diagnostics for type A static correlation[38] (a.k.a. left−right static correlation,[39] absolute near-degeneracy correlation): D$_1$ [defined as[40] $\lambda_{max}(\mathbf{T}_1 \cdot \mathbf{T}_1^\dagger)^{1/2}$ where $\mathbf{T}_1$ is the single excitations amplitude vector], $1 - C_0^2$ (i.e., one minus the squared coefficient of the reference determinant in a CASSCF calculation with an appropriate active space), the $M$ diagnostic proposed by Truhlar and co-workers[41] (which for closed-shell systems reduces to $1 - n_{HOMO}/[2 + (n_{LUMO}/2)]$), and Matito's $I_{ND}$ diagnostic[42] based on natural orbital occupations. A fairly recent review of static correlation diagnostics can be found in ref 43. Additional diagnostics for the considered systems are discussed in ref 6. As shown there, FOD analysis[44] (as expected) indicates that static correlation is smeared out over the entire molecule. (As an aside, using the same (12,12) active space in all CASSCF calculations for the purpose of calculating $1 - C_0^2$ sidesteps the issue pointed out by a reviewer that $1 - C_0^2$ is not size-extensive. A workaround to bring $1 - C_0^2$ on the same scale for different numbers of correlated electrons was pointed out in endnote 31 of the work of Via-Nadal et al.:[45] tt effectively amounts to replacing $x \equiv 1 - C_0^2$ by $1 - (1-x)^{1/a}$ where $a \equiv N_{val}/N_{val,ref}$ or the number of valence electrons divided by the number for a reference system. If $x$ is not too large, MacLaurin expansion in $x$ yields $1 - (1-x)^{1/a} = x/a + (a-1)x^2/2a^2 + ... \approx x/a$ and hence $1 - (1-x)^{1/a} \approx (1 - C_0^2)N_{val,ref}/N_{val}$.

Finally, in order to verify that SCF orbitals for all structures in all codes correspond to global minima on the $S = 0$ Hartree−Fock energy hypersurface, relative energies at the SCF level were compared to those obtained in the same basis set using stable = (int,opt) in Gaussian 16[46] and found to agree to 0.01 kcal mol$^{-1}$ or better. Some of the Möbius structures, in particular, required care to ensure convergence to the correct state with the other codes: in the most refractory case, **32M$_{2b}$**, we resorted to HF/STO-3G on the quadruple cation, used as initial guess for HF/STO-3G on the neutral species, finally used in turn as initial guess for HF/cc-pVDZ.

## RESULTS AND DISCUSSION

**Adequacy of the Canonical Reference Level.** As mentioned in the Introduction, the largest basis set for which we were able to obtain fully canonical CCSD(T) answers for comparison was the cc-pVDZ (no p on hydrogen) basis set.[47] The mind wonders whether, at least for the problem at hand, this level of theory is sufficiently close to the FCI/CBS (full configuration interaction/complete basis set) limit to be adequate as a canonical reference point.

Concerning the first aspect, i.e., post-CCSD(T) correlation effects, the size of the system clearly precludes carrying out CCSDT(Q) let alone CCSDTQ calculations. However, for limited orbital active spaces, we were able to carry out ICE-CI (iterative configuration expansion—configuration interaction—ICE-CI is effectively ORCA's implementation of Malrieu's CIPSI algorithm[48]) calculations using ORCA and compare them to CCSD(T) in the same orbital space. The result, for active spaces ranging from 12-electrons-in-12-orbitals, or (12,12) for short, to (30,30) are given in Table 1. Clearly, at least for the property of interest, post-CCSD(T) corrections are surprisingly small. This may, of course, be the result of a fortunate error compensation between neglect of higher-order iterative triple substitution effects CCSDT-CCSD(T) and neglect of connected quadruple excitations. Similar cancellations are seen in the atomization energies of some small molecules with multireference character, e.g., $C_2$.[49−51]

Concerning the second aspect, i.e., basis set incompleteness, we were able to carry out canonical explicitly correlated[52,53] RI-MP2-F12 calculations with the cc-pVDZ-F12 basis set[54] and associated auxiliary basis sets[55] for all species. For the largest macrocycle (i.e., the [32]heptaphyrin), such calculations required about 10TB of scratch space each, which we "jury-rigged" by cross-mounting SSD scratch directories from other nodes through NFS-over-InfiniBand. Typically (see, e.g., reviews on F12 theory[52,53]), F12 calculations with appropriate basis sets gain about 2−3 "zetas" in basis set convergence. Hence, the MP2-F12/cc-pVDZ-F12 energetics ought to be comparable or superior to MP2/cc-pVQZ in terms of convergence.

We can easily verify this in the present context, of course, by carrying out RI-MP2/cc-pVTZ and cc-pVQZ calculations and extrapolating to the complete basis set limit using the Helgaker formula.[56] In this event, MP2/cc-pV{T,Q}Z relative energies thus obtained deviate from their MP2-F12/cc-pVDZ-F12 counterparts by less than 0.1 kcal mol$^{-1}$ RMSD. The basis set extension effect itself, from MP2/cc-pVDZ, is just 0.9 kcal mol$^{-1}$ RMSD in both cases. We may thus safely assume that the coupling term C in the equation below is negligible:

$$\text{CCSD(T)/LARGE} = \text{CCSD(T)/SMALL} + \text{MP2/LARGE} - \text{MP2/SMALL} + C \quad (1)$$

$$C = [\text{CCSD(T)-MP2}]/\text{LARGE} - [\text{CCSD(T)-MP2}]/\text{SMALL} \quad (2)$$

and thus, that we can make the familiar "high-level correction" (HLC) approximation:

$$\text{CCSD(T)/LARGE}$$
$$\approx [\text{CCSD(T)-MP2}]/\text{SMALL} + \text{MP2/LARGE}$$
$$= \text{HLC/SMALL} + \text{MP2/LARGE} \quad (3)$$

(For a discussion of one-particle/"basis set" vs n-particle space/"electron correlation method" coupling, see ref 57.) Our best estimates for the relative energies of the topology interconversions in our expanded porphyrin database are collected in Table 2. For the purpose of assessing localized methods against canonical results, however, the above gives us confidence that CCSD(T)/cc-pVDZ is a reasonable starting point.

Table 2. Our Best Estimates for the Relative Isomer Energies Considered in This Work[a]

| system | CCSD(T)/cc-pVDZ | MP2/cc-pV{T,Q}Z + [CCSD(T)-MP2]/cc-pVDZ | MP2-F12/cc-pVDZ-F12 + [CCSD(T)-MP2]/cc-pVDZ |
|---|---|---|---|
| 24H$_a$ | 9.12 | 7.92 | 8.06 |
| 24H$_b$ | 0.00 | 0.00 | 0.00 |
| 24M | 6.06 | 6.38 | 6.48 |
| 24TS$_1$ | 9.05 | 8.93 | 9.01 |
| 24TS$_2$ | 4.87 | 5.12 | 5.18 |
| 28H | 0.00 | 0.00 | 0.00 |
| 28M | −0.73 | −1.77 | −1.75 |
| 28M$_{1A}$ | 0.46 | 0.28 | 0.28 |
| 28M$_{1B}$ | 1.82 | 1.39 | 1.39 |
| 28F | −0.38 | 0.16 | −0.08 |
| 28TS$_{1A}$ | 6.33 | 4.65 | 4.58 |
| 28TS$_{1B}$ | 2.86 | 2.00 | 1.92 |
| 28TS$_{2A}$ | 6.87 | 6.10 | 6.02 |
| 28TS$_{2B}$ | 9.89 | 8.88 | 8.79 |
| 28TS$_3$ | 5.17 | 4.50 | 4.36 |
| 32F | 0.00 | 0.00 | 0.00 |
| 32M$_a$ | 16.81 | 15.45 | 15.65 |
| 32M$_b$ | 16.74 | 16.59 | 16.52 |
| 32H | 34.60 | 32.59 | 32.72 |
| 32TS$_1$ | 17.49 | 16.08 | 16.16 |
| 32TS$_2$ | 33.79 | 32.33 | 32.36 |

[a]The latter were obtained at the MP2/cc-pV{T,Q}Z + [CCSD(T)-MP2]/cc-pVDZ and MP2-F12/cc-pVDZ-F12 + [CCSD(T)-MP2]/cc-pVDZ levels of theory, with p functions on H omitted. All entries are in kcal mol$^{-1}$.

**Initial Assessment of the Localized vs Canonical Methods.** For heptaphyrin, each canonical CCSD(T) required about a week on eight 16-core Intel Haswell nodes, with MOLPRO running a 3-level parallelism of nodes, processes, and [in (T) and LAPACK] OpenMP threads. *In contrast, the corresponding localized calculations took from a few hours to 1 day on just a single node.* A comparison of various approximate PNO-CCSD(T) relative energies with the canonical reference values is given in Table 3, and the box-and-whisker plots for different localized approaches are shown in Figure 2.

First of all, DLPNO-CCSD(T$_1$) with tight PNO settings appears to be the overall best performer among all PNO-type approaches, having an RMSD of only 1.43 kcal mol$^{-1}$ from the reference. Resorting to default PNO settings raises the error by only ∼0.4 kcal mol$^{-1}$, while reducing wall time by about 75−80%, and may therefore be a desirable option in cases where tight PNO settings become too computationally demanding.

Table 3. Canonical CCSD(T) Relative Energies (kcal mol$^{-1}$) and Errors with Various Localized Orbital CCSD(T) Approximations for the Relative Energies of [24]N-Fused Pentaphyrin, [28]Hexaphyrin, and [32]Heptaphyrin Structures (F = figure-eight, M = Möbius, H = Hückel; TS = transition states)[a]

| System | CCSD(T) cc-pVDZ | DLPNO-CCSD(T$_0$) ORCA | | DLPNO-CCSD(T$_1$) ORCA | | PNO-LCCSD(T$_1$) MOLPRO | | LNO-CCSD(T) MRCC | | | Static correlation diagnostics[i] | | | |
|---|---|---|---|---|---|---|---|---|---|---|---|---|---|---|
| | canonical | Normal[b] | Tight[c] | Normal[b] | Tight[c] | Normal[d] | Tight[e] | Normal[f] | Tight[g] | Tight+[h] | $D_1$ | $(1-C_0^2)$ | $M_{diag}$ | $I_{ND}$ |
| 24H$_a$ | 9.1 | 0.8 | 0.3 | 0.8 | 0.2 | 0.3 | 0.3 | 0.6 | 0.1 | 0.1 | 0.081 | 0.117 | 0.096 | 0.118 |
| 24H$_b$ | 0.0 | 0.0 | 0.0 | 0.0 | 0.0 | 0.0 | 0.0 | 0.0 | 0.0 | 0.0 | 0.079 | 0.122 | 0.094 | 0.111 |
| 24M | 6.1 | 0.5 | 0.7 | 0.4 | 0.4 | 0.6 | 0.6 | 0.2 | 0.0 | 0.0 | 0.088 | 0.141 | 0.112 | 0.130 |
| 24TS$_1$ | 9.0 | -0.1 | -0.1 | -0.1 | -0.1 | -0.2 | -0.2 | 0.0 | 0.1 | 0.1 | 0.078 | 0.129 | 0.097 | 0.115 |
| 24TS$_2$ | 4.9 | 0.1 | 0.2 | 0.0 | 0.1 | 0.2 | 0.2 | 0.0 | 0.0 | 0.0 | 0.086 | 0.132 | 0.102 | 0.118 |
| 28F | -0.4 | -1.4 | -1.0 | -1.5 | -1.0 | -0.7 | -0.6 | -1.5 | -1.1 | -0.6 | 0.077 | 0.132 | 0.094 | 0.145 |
| 28M$_{1A}$ | 0.5 | 2.1 | 3.0 | 1.3 | 1.7 | 5.7 | 4.2 | 0.8 | 0.0 | -0.4 | 0.103 | 0.192 | 0.165 | 0.203 |
| 28M | -0.7 | 2.8 | 2.9 | 2.1 | 1.7 | 5.1 | 3.7 | 0.4 | -0.1 | -0.3 | 0.108 | 0.183 | 0.153 | 0.196 |
| 28M$_{1B}$ | 1.8 | 3.0 | 2.9 | 2.5 | 1.9 | 6.1 | 4.5 | 0.2 | 0.0 | -0.4 | 0.110 | 0.193 | 0.165 | 0.203 |
| 28TS$_3$ | 5.2 | -1.0 | -0.5 | -0.9 | -0.4 | -0.4 | -0.2 | -1.4 | -1.0 | -0.5 | 0.092 | 0.129 | 0.096 | 0.141 |
| 28H | 0.0 | 0.0 | 0.0 | 0.0 | 0.0 | 0.0 | 0.0 | 0.0 | 0.0 | 0.0 | 0.081 | 0.141 | 0.107 | 0.155 |
| 28TS$_{1A}$ | 6.3 | -0.4 | 0.0 | -0.3 | 0.1 | 0.2 | 0.1 | -0.4 | -0.5 | -0.1 | 0.095 | 0.130 | 0.101 | 0.138 |
| 28TS$_{1B}$ | 2.9 | -0.8 | -0.5 | -0.8 | -0.4 | -0.4 | -0.3 | -0.9 | -0.6 | -0.3 | 0.082 | 0.137 | 0.104 | 0.145 |
| 28TS$_{2A}$ | 6.9 | 3.1 | 3.7 | 1.8 | 2.2 | 6.8 | 4.8 | 0.7 | 0.0 | -0.3 | 0.115 | 0.186 | 0.156 | 0.197 |
| 28TS$_{2B}$ | 9.9 | 3.1 | 3.3 | 2.3 | 2.0 | 6.0 | 4.3 | 0.3 | 0.0 | -0.3 | 0.116 | 0.183 | 0.153 | 0.194 |
| 32F | 0.0 | 0.0 | 0.0 | 0.0 | 0.0 | 0.0 | 0.0 | 0.0 | 0.0 | 0.0 | 0.088 | 0.146 | 0.098 | 0.161 |
| 32H | 34.6 | 0.0 | -0.2 | 0.3 | 0.3 | -2.2 | -0.9 | 1.6 | 1.0 | 0.7 | 0.084 | 0.137 | 0.096 | 0.150 |
| 32TS$_2$ | 33.8 | -0.5 | -0.4 | -0.1 | 0.1 | -2.6 | -1.2 | 1.2 | 0.7 | 0.5 | 0.084 | 0.128 | 0.098 | 0.146 |
| 32M$_a$ | 16.8 | 4.5 | 3.4 | 4.0 | 2.5 | 4.0 | 3.5 | 3.9 | 1.7 | 0.8 | 0.117 | 0.188 | 0.156 | 0.207 |
| 32M$_b$ | 16.7 | 5.2 | 4.2 | 4.7 | 3.3 | 5.7 | 5.0 | 3.5 | 1.6 | 0.7 | 0.131 | 0.196 | 0.170 | 0.216 |
| 32TS$_1$ | 17.5 | 0.0 | -0.3 | 0.1 | -0.1 | -1.2 | -0.5 | 0.8 | 0.2 | 0.3 | 0.096 | 0.132 | 0.102 | 0.154 |
| RMSD | REFERENCE | 2.27 | 2.14 | 1.88 | 1.43 | 3.66 | 2.73 | 1.47 | 0.74 | 0.42 | | | | |
| Möbius (-like) | | 3.32 | 3.17 | 2.73 | 2.11 | 5.33 | 4.04 | 1.90 | 0.82 | 0.46 | | | | |
| Other structures | | 0.69 | 0.43 | 0.67 | 0.37 | 1.20 | 0.55 | 1.01 | 0.66 | 0.38 | | | | |

[a]RMSDs from canonical results in the same basis set (kcal mol$^{-1}$). Energy differences are heat-mapped on a continuous gradient from blue (most negative value) via white (zero) to red (most positive value); diagnostics are heat-mapped green (low) via orange to red (high) on a continuous percentile gradient for each column. [b]NormalPNO. [c]tightPNO. [d]defaultDomain. [e]tightDomain. [f]lcorthr = normal. [g]lcorthr = tight. [h]lcorthr = tight, wpairtol = 1 × 10$^{-6}$. [i]Taken from Table 2 in ref 6. $(1 - C_0^2)$ was obtained at the CASSCF(12,12) level for all species, M and $I_{ND}$ at the ICE-FCI(30/30) level, and $D_1$ at the CCSD/cc-pVDZ(no $p$ functions on H) level.

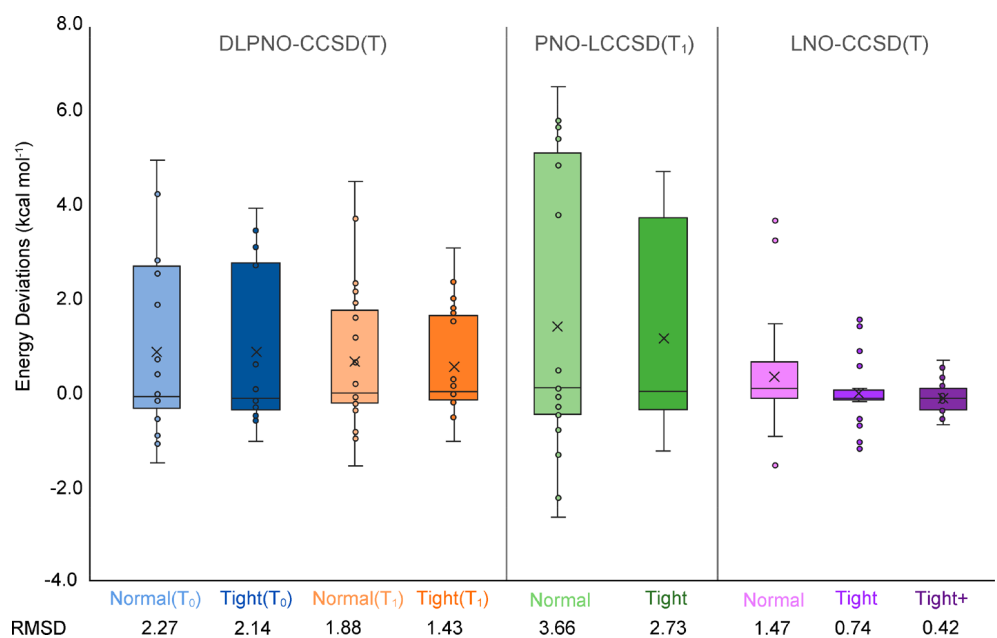

**Figure 2.** Box-and-whisker plots for various localized orbital CCSD(T) approximations, showing the error spread for the expanded porphyrin database with respect to canonical CCSD(T) energies. The RMSDs (in kcal mol$^{-1}$) are also displayed below each method.

Table 4. Canonical MP2 Relative Energies (kcal mol$^{-1}$) and Errors with Various Localized Orbital MP2 Approximations for the Relative Energies of the Expanded Porphyrins under Consideration[a]

| Basis: cc-pVDZ, p functions on H omitted | MP2 | DLPNO-MP2 | | PNO-LMP2 | | LMP2 of MRCC | | |
|---|---|---|---|---|---|---|---|---|
| | Canonical | Normal[b] | Tight[c] | Normal[d] | Tight[e] | Normal[f] | Tight[g] | Tight+[h] |
| 24H$_a$ | 8.5 | -0.1 | 0.0 | -0.5 | 0.0 | 0.4 | 0.1 | 0.0 |
| 24H$_b$ | 0.0 | 0.0 | 0.0 | 0.0 | 0.0 | 0.0 | 0.0 | 0.0 |
| 24M | 3.2 | 0.0 | 0.0 | 0.2 | 0.2 | 0.3 | 0.0 | 0.0 |
| 24TS$_1$ | 9.5 | -0.1 | 0.0 | -0.6 | -0.1 | -0.1 | 0.0 | 0.0 |
| 24TS$_2$ | 4.2 | 0.0 | 0.0 | -0.1 | 0.1 | 0.1 | 0.0 | 0.0 |
| 28F | -2.3 | 0.0 | -0.1 | -0.4 | -0.3 | -1.0 | -0.8 | -0.3 |
| 28M$_{1A}$ | -18.8 | 0.5 | 0.2 | 4.7 | 2.2 | 1.5 | 0.4 | 0.1 |
| 28M | -14.6 | 0.5 | 0.2 | 3.3 | 1.6 | 0.8 | 0.2 | 0.0 |
| 28M$_{1B}$ | -16.3 | 0.4 | 0.2 | 4.5 | 2.2 | 1.4 | 0.5 | 0.1 |
| 28TS$_3$ | 6.0 | 0.0 | 0.0 | -0.3 | -0.2 | -1.0 | -0.8 | -0.3 |
| 28R | 0.0 | 0.0 | 0.0 | 0.0 | 0.0 | 0.0 | 0.0 | 0.0 |
| 28TS$_{1A}$ | 8.5 | -0.1 | -0.1 | -0.1 | 0.0 | -0.4 | -0.5 | -0.2 |
| 28TS$_{1B}$ | 3.3 | 0.0 | 0.0 | -0.3 | -0.1 | -0.7 | -0.5 | -0.1 |
| 28TS$_{2A}$ | -13.3 | 0.6 | 0.3 | 5.2 | 2.4 | 1.2 | 0.2 | 0.0 |
| 28TS$_{2B}$ | -7.6 | 0.5 | 0.2 | 3.9 | 1.7 | 0.8 | 0.2 | 0.0 |
| 32F | 0.0 | 0.0 | 0.0 | 0.0 | 0.0 | 0.0 | 0.0 | 0.0 |
| 32H | 40.8 | -0.6 | -0.2 | -1.8 | -0.6 | 0.5 | 0.6 | 0.4 |
| 32TS$_2$ | 41.0 | -0.5 | -0.1 | -2.1 | -0.7 | 0.2 | 0.3 | 0.2 |
| 32M$_a$ | 10.2 | 0.0 | 0.1 | 1.9 | 1.1 | 2.6 | 1.1 | 0.6 |
| 32M$_b$ | 5.0 | 0.1 | 0.1 | 3.1 | 1.9 | 2.4 | 1.3 | 0.6 |
| 32TS$_1$ | 21.1 | -0.3 | -0.2 | -1.0 | -0.4 | 0.1 | 0.0 | 0.1 |
| RMSD | REF | 0.33 | 0.14 | 2.56 | 1.22 | 1.12 | 0.57 | 0.26 |
| Möbius(-like) | | 0.39 | 0.18 | 3.68 | 1.78 | 1.57 | 0.66 | 0.31 |
| Others | | 0.27 | 0.09 | 0.98 | 0.34 | 0.56 | 0.48 | 0.20 |

[a]RMSDs from canonical results in the same basis set likewise in kcal mol$^{-1}$. Energy differences are heat-mapped on a continuous gradient from blue (most negative value) via white (zero) to red (most positive value). [b]NormalPNO (ORCA). [c]tightPNO (ORCA). [d]defaultDomain (MOLPRO). [e]tightDomain (MOLPRO). [f]lcorthr = normal (MRCC). [g]lcorthr = tight. [h]lcorthr = tight, wpairtol = 1 × 10$^{-6}$.

DLPNO-CCSD(T$_0$) does not measure up to the former scheme—exhibiting 2.14 and 2.27 kcal mol$^{-1}$ RMSDs from the canonical energies using tight and default PNO settings, respectively. Indeed, the difference associated with the latter settings is not as large as in the (T$_1$) case—the domain improvement "drowns in the noise" of the T$_0$ approximation, so to speak.

PNO-LCCSD(T$_1$) seemingly offers the least-satisfactory performance among this class of localized methods, deviating from the reference values by 3.66 and 2.73 kcal mol$^{-1}$ using default and tight PNO settings, respectively. The latter PNO settings are clearly superior in this case.

LNO-CCSD(T) performs exceptionally well compared to the above PNO-type approaches—having an RMSD of 1.47 kcal mol$^{-1}$ with default settings (lcorthr = Normal) and just 0.74 kcal mol$^{-1}$ with (lcorthr = Tight). Further tightening thresholds to lcorthr = vtight approximately quadruples the total CPU time; fortunately, we found that just tightening wpairtol from 3 × 10$^{-6}$ to 1 × 10$^{-6}$, while leaving the remaining parameters at their lcorthr = Tight values, recovered most of the improvement at minimal additional cost. This setting is denoted as "Tight+" in Tables 3–7. RMSD from canonical CCSD(T) for Tight+ was found to be just 0.42 kcal mol$^{-1}$.

As can be seen in Table 3 and Figure 3 for the expanded porphyrin database, deviations of DLPNO and PNO from canonical relative energies are found to be statistically correlated with several diagnostics for the type A static correlation (i.e., absolute near-degeneracy). Indeed, the largest values for all four diagnostics, on the one hand, and the largest deviations from canonical energetics, on the other hand, are specifically observed for the Möbius structures and for two Möbius-like transition states (28TS$_{2A}$ and 28TS$_{2B}$). The M diagnostic, in particular, turns out to be a fair predictor for the energy difference between the localized orbital approaches and canonical CCSD(T) method, with $R^2 = 0.89$ for DLPNO-CCSD(T$_1$) (Figure 3a) and $R^2 = 0.96$ for PNO-LCCSD(T$_1$) (Figure 3b), both of them with Tight settings.

We found it informative, then, to break down error statistics between Möbius(-like) structures vs the Hückel and twisted-Hückel structures. At the bottom of Table 3, we then see that, for the non-Möbius structures, all three PNO methods can reach about 0.5 kcal mol$^{-1}$ RMSD on Tight settings and about 1 kcal mol$^{-1}$ on regular settings. However, for the Möbius structures, much more pronounced errors are seen.

A similar discrepancy can be observed for the DLPNO methods. Indeed, for the non-Möbius structures, RMSD is just 0.7 kcal mol$^{-1}$ for NormalPNO and 0.4 kcal mol$^{-1}$ for TightPNO, while these figures jump up to 2 kcal mol$^{-1}$ for DLPNO-CCSD(T$_1$) with TightPNO and to 3 kcal mol$^{-1}$ for the other options, when considering only the Möbius-type structures. Also, while T$_0$ and T$_1$ are essentially indistinguish-able in quality for the non-Möbius systems, T$_1$ is markedly superior for the Möbius ones.

**Table 5. Canonical CCSD Relative Energies (kcal mol$^{-1}$) and Errors with Various Localized Orbital CCSD Approximations for the Relative Energies of the Expanded Porphyrins under Consideration**[a]

| cc-pVDZ no p on H | CCSD canonical | differences from canonical CCSD | | | | | | |
|---|---|---|---|---|---|---|---|---|
| | | DLPNO-CCSD | | PNO-LCCSD | | LNO-CCSD | | |
| | | Normal[b] | Tight[c] | Normal[d] | Tight[e] | Normal[f] | Tight[g] | Tight+[h] |
| 24H$_a$ | 9.2 | 0.3 | 0.0 | -0.4 | 0.1 | 0.2 | 0.0 | 0.0 |
| 24H$_b$ | 0.0 | 0.0 | 0.0 | 0.0 | 0.0 | 0.0 | 0.0 | 0.0 |
| 24M | 8.1 | -0.4 | 0.1 | 0.2 | 0.3 | 0.3 | 0.1 | 0.1 |
| 24TS$_1$ | 8.7 | 0.1 | 0.0 | -0.6 | -0.1 | 0.0 | 0.1 | 0.1 |
| 24TS$_2$ | 5.6 | -0.2 | 0.0 | 0.0 | 0.1 | 0.1 | 0.1 | 0.1 |
| 28F | 0.4 | -0.9 | -0.6 | -0.4 | -0.2 | -0.6 | -0.5 | -0.1 |
| 28M$_{1A}$ | 9.7 | -2.1 | 0.5 | 3.0 | 2.0 | 1.0 | 0.3 | 0.1 |
| 28M | 7.0 | -0.9 | 0.7 | 2.7 | 1.8 | 0.6 | 0.2 | 0.2 |
| 28M$_{1B}$ | 11.1 | -1.1 | 0.5 | 3.2 | 2.2 | 0.7 | 0.2 | 0.1 |
| 28TS$_3$ | 5.7 | -0.3 | -0.1 | -0.1 | 0.0 | -0.5 | -0.4 | 0.0 |
| 28R | 0.0 | 0.0 | 0.0 | 0.0 | 0.0 | 0.0 | 0.0 | 0.0 |
| 28TS$_{1A}$ | 6.6 | 0.1 | 0.3 | 0.2 | 0.2 | -0.1 | -0.3 | 0.0 |
| 28TS$_{1B}$ | 2.8 | -0.2 | -0.1 | -0.2 | -0.1 | -0.4 | -0.4 | -0.1 |
| 28TS$_{2A}$ | 17.4 | -1.8 | 1.0 | 3.7 | 2.4 | 0.7 | 0.2 | 0.1 |
| 28TS$_{2B}$ | 19.8 | -1.1 | 0.8 | 3.1 | 2.0 | 0.5 | 0.2 | 0.2 |
| 32F | 0.0 | 0.0 | 0.0 | 0.0 | 0.0 | 0.0 | 0.0 | 0.0 |
| 32H | 31.4 | 1.8 | 0.5 | -1.6 | -0.7 | 0.1 | 0.4 | 0.2 |
| 32TS$_2$ | 30.2 | 1.7 | 0.5 | -1.8 | -0.7 | 0.0 | 0.3 | 0.1 |
| 32M$_a$ | 22.0 | 1.2 | 0.8 | 1.6 | 1.4 | 2.1 | 1.2 | 0.8 |
| 32M$_b$ | 24.1 | 1.0 | 1.2 | 2.5 | 2.2 | 1.7 | 1.0 | 0.6 |
| 32TS$_1$ | 15.8 | 1.3 | 0.1 | -0.9 | -0.3 | 0.2 | 0.1 | 0.2 |
| RMSD | | 1.12 | 0.57 | 1.92 | 1.29 | 0.78 | 0.46 | 0.27 |
| Möbius(-like) | REFERENCE | 1.30 | 0.78 | 2.72 | 1.89 | 1.13 | 0.60 | 0.39 |
| Other structures | | 0.95 | 0.32 | 0.86 | 0.35 | 0.30 | 0.31 | 0.12 |

[a]RMSDs from canonical results in the same basis set likewise in kcal mol$^{-1}$. Energy differences are heat-mapped on a continuous gradient from blue (most negative value) via white (zero) to red (most positive value). [b]NormalPNO (ORCA). [c]tightPNO (ORCA). [d]defaultDomain (MOLPRO). [e]tightDomain (MOLPRO). [f]lcorthr = normal (MRCC). [g]lcorthr = tight. [h]lcorthr = tight, wpairtol = 1 × 10$^{-6}$.

The deficiencies of the T$_0$ approximation for systems with some static correlation are of course not unique to the macrocycles at hand. In the original DLPNO-CCSD(T$_1$) paper,[29] it was shown that, for small-gap systems, the (T$_0$) approximation breaks down and relative energies show substantial deviations from the parent canonical CCSD(T) results. Relatedly, we point to the work of Iron and Janes on metal–organic barrier heights (MOBH35),[58,59] where a comparatively small, yet significant, difference of almost 1 kcal mol$^{-1}$ RMSD was found between DLPNO-CCSD(T$_0$) and DLPNO-CCSD(T$_1$) barrier heights.[59] Efremenko and Martin[60] found more significant differences for the mechanisms of Ru(II) and Ru(III) catalyzed hydroarylation and oxidative coupling.[61]

While the Möbius RMSD does get worse from (T$_1$) to (T$_0$), it is a difference of degree and not of kind. Switching from "Normal" to "Tight" criteria actually has the largest impact for LNO-CCSD(T), where it cuts the remaining error for the Möbius structures by over half; a significant improvement is also seen for DLPNO-CCSD(T$_1$). In stark contrast, while the LNO-CCSD(T) approach with Normal criteria does exhibit nearly twice the RMSD (1.9 kcal mol$^{-1}$) for Möbius as for non-Möbius (1.0 kcal mol$^{-1}$), for Tight and especially Tight+ criteria, this difference essentially vanishes. Indeed, for Tight+ all errors are below 1 kcal mol$^{-1}$.

**Component Breakdown of Localized vs Canonical Methods.** Let us now decompose the relative canonical CCSD(T) energies into their MP2 and CCSD building blocks, in order to get deeper insights regarding the relationship between the canonical and PNO-based methods considered above.

As can be seen in Table 4, PNO-LMP2 with default PNO settings performs rather poorly, having a RMSD of no less than 2.6 kcal mol$^{-1}$ from the canonical reference values. Resorting to tight PNO domains does lead to an improvement, reducing the error by more than half (1.2 kcal mol$^{-1}$ RMSD). Like for the complete CCSD(T) energetics, it can be seen that the Möbius structures are responsible for most of the observed errors—RMSD = 3.7 (Normal) and 1.8 (Tight) kcal mol$^{-1}$ versus, for the non-Möbius structures, 1.0 and 0.3 kcal mol$^{-1}$, respectively.

ORCA's DLPNO-MP2, on the other hand, has no such large discrepancy between Möbius and non-Möbius systems. Overall RMSD = 0.33 kcal mol$^{-1}$ (NormalPNO) and 0.14 kcal mol$^{-1}$ (TightPNO), the latter functionally equivalent in quality to the reference values.

For LNO-MP2, we do see a substantial difference between Möbius and non-Möbius structures at default settings, but this is much reduced for Tight and especially Tight+ settings. Overall, RMSD for Tight settings (0.57 kcal mol$^{-1}$) is still almost double that for NormalPNO DLPNO-MP2 (0.33 kcal mol$^{-1}$), while Tight+ slightly reduces the statistical errors to RMSD = 0.26 kcal mol$^{-1}$.

We shall now move on to the CCSD contributions (Table 5). For LNO-CCSD(T), we have followed the recommendation from Nagy et al.[35] to split the weak-pair MP2 corrections evenly between CCSD and (T). It can be seen that DLPNO-CCSD gets closer to canonical CCSD in the same basis set compared to PNO-LCCSD; even DLPNO-CCSD with DefaultPNO settings, at RMSD = 1.1 kcal mol$^{-1}$, outperforms PNO-LCCSD with TightDomain settings (1.3 kcal mol$^{-1}$), and with default domain settings, RMSD for PNO-LCCSD even increases to 1.9 kcal mol$^{-1}$. For the non-Möbius systems, DLPNO-CCSD and PNO-LCCSD are virtually indistinguishable in performance; for the Möbius structures, PNO-LCCSD exhibits more significant errors (2.7 and 1.9 kcal mol$^{-1}$ RMSD for default and tight settings, respectively), whereas DLPNO-CCSD seems to offer a more satisfying performance (1.3 and 0.8 kcal mol$^{-1}$).

LNO-CCSD with Tight+ settings, at RMSD = 0.27 kcal mol$^{-1}$, is the best performer; RMSD climbs to 0.46 kcal mol$^{-1}$ for tight settings, still superior to DLPNO-CCSD TightPNO. Here too, we observe a performance difference between Möbius and non-Möbius structures.

What about the (T) contribution when considered in isolation? As we have seen for the full CCSD(T) relative energies, there is little difference between DLPNO-(T$_0$) and DLPNO-(T$_1$) for the non-Möbius structures, and this applies for both NormalPNO and TightPNO. Nevertheless, for the Möbius structures, the difference is quite pronounced with DLPNO-(T$_1$) TightPNO having only about one-half the error (0.9 kcal mol$^{-1}$) of DLPNO-(T$_0$) TightPNO and about one-third the error of DLPNO-(T$_0$) NormalPNO. LNO-CCSD(T) can, however, match the best result even with normal settings, while on TightPNO RMSD drops to just 0.3 kcal mol$^{-1}$, with no noticeable difference between Möbius and non-Möbius.

Our attempts to carry out PNO-LCCSD(T)-F12/cc-pVDZ-F12 calculations[30,62] on these extended π-systems met with failure for technical reasons. Presumably, if we were able to run them to completion, they would be much closer to the canonical basis set limit than PNO-LCCSD(T) is to its canonical counterpart.

Table 6. Canonical (T) Relative Energies (kcal mol$^{-1}$) and Errors with Various Localized Orbital (T) Approximations for the Relative Energies of the Expanded Porphyrins under Consideration[a]

| cc-pVDZ no p on H | CCSD(T) | differences from canonical (T) | | | | | | | |
|---|---|---|---|---|---|---|---|---|---|
| | | DLPNO-CCSD($T_0$) | | DLPNO-CCSD($T_1$) | | PNO-LCCSD($T_1$) | | LNO-CCSD(T) | |
| | canonical | Normal[b] | Tight[c] | Normal[b] | Tight[c] | Normal[d] | Tight[e] | Normal[f] | Tight[g] |
| $24H_a$ | -0.1 | 0.5 | 0.2 | 0.5 | 0.1 | 0.6 | 0.2 | 0.4 | 0.2 |
| $24H_b$ | 0.0 | 0.0 | 0.0 | 0.0 | 0.0 | 0.0 | 0.0 | 0.0 | 0.0 |
| 24M | -2.1 | 0.9 | 0.6 | 0.7 | 0.3 | 0.4 | 0.3 | -0.1 | -0.1 |
| $24TS_1$ | 0.3 | -0.1 | 0.0 | -0.2 | 0.0 | 0.4 | -0.1 | 0.0 | 0.0 |
| $24TS_2$ | -0.7 | 0.3 | 0.2 | 0.2 | 0.1 | 0.3 | 0.1 | -0.1 | -0.1 |
| 28F | -0.7 | -0.5 | -0.4 | -0.6 | -0.4 | -0.4 | -0.4 | -0.9 | -0.6 |
| $28M_{1A}$ | -9.3 | 4.2 | 2.5 | 3.4 | 1.2 | 2.7 | 2.2 | -0.2 | -0.3 |
| 28M | -7.8 | 3.7 | 2.2 | 3.0 | 1.0 | 2.4 | 1.9 | -0.3 | -0.3 |
| $28M_{1B}$ | -9.3 | 4.2 | 2.4 | 3.7 | 1.3 | 2.9 | 2.3 | -0.5 | -0.2 |
| $28TS_3$ | -0.5 | -0.7 | -0.4 | -0.6 | -0.3 | -0.2 | -0.2 | -0.9 | -0.6 |
| 28R | 0.0 | 0.0 | 0.0 | 0.0 | 0.0 | 0.0 | 0.0 | 0.0 | 0.0 |
| $28TS_{1A}$ | -0.3 | -0.5 | -0.3 | -0.4 | -0.2 | 0.0 | 0.0 | -0.3 | -0.2 |
| $28TS_{1B}$ | 0.1 | -0.6 | -0.4 | -0.5 | -0.3 | -0.2 | -0.2 | -0.5 | -0.2 |
| $28TS_{2A}$ | -10.6 | 4.8 | 2.7 | 3.5 | 1.2 | 3.1 | 2.4 | 0.0 | -0.1 |
| $28TS_{2B}$ | -9.9 | 4.2 | 2.5 | 3.4 | 1.2 | 2.8 | 2.2 | -0.2 | -0.2 |
| 32F | 0.0 | 0.0 | 0.0 | 0.0 | 0.0 | 0.0 | 0.0 | 0.0 | 0.0 |
| 32H | 3.2 | -1.8 | -0.7 | -1.5 | -0.3 | -0.6 | -0.2 | 1.5 | 0.6 |
| $32TS_2$ | 3.6 | -2.2 | -0.9 | -1.9 | -0.4 | -0.8 | -0.4 | 1.3 | 0.4 |
| $32M_a$ | -5.2 | 3.3 | 2.5 | 2.8 | 1.7 | 2.4 | 2.1 | 1.8 | 0.5 |
| $32M_b$ | -7.4 | 4.2 | 3.0 | 3.7 | 2.1 | 3.2 | 2.8 | 1.8 | 0.5 |
| $32TS_1$ | 1.6 | -1.3 | -0.4 | -1.2 | -0.2 | -0.4 | -0.2 | 0.6 | 0.1 |
| RMSD | | 2.69 | 1.64 | 2.23 | 0.92 | 1.78 | 1.45 | 0.85 | 0.34 |
| Möbius(-like) | REFERENCE | 3.86 | 2.40 | 3.18 | 1.34 | 2.63 | 2.16 | 0.92 | 0.32 |
| Other structures | | 1.07 | 0.48 | 0.93 | 0.27 | 0.45 | 0.23 | 0.79 | 0.36 |

[a]RMSDs from canonical results in the same basis set likewise in kcal mol$^{-1}$. Energy differences are heat-mapped on a continuous gradient from blue (most negative value) via white (zero) to red (most positive value. [b]NormalPNO (ORCA). [c]tightPNO (ORCA). [d]defaultDomain (MOLPRO). [e]tightDomain (MOLPRO). [f]Normal settings (MRCC). [g]Tight settings (MRCC).

This comparatively weak basis set sensitivity beyond cc-pVDZ (Table 2) indicates that thermodynamic equilibria in the present systems are primarily driven by nondynamical correlation effects—which are well-known (e.g., ref 49) to converge fairly rapidly with the basis set—rather than the slowly converging dynamical correlation contributions. In such a scenario, especially for still larger systems like octaphyrins or decaphyrins, it may be attractive not just to combine MP2 in a large basis set with a "high-level correction", i.e., the aggregate post-MP2 correction [CCSD(T)-MP2] from a small basis set, but to obtain the latter using a DLPNO, PNO-L, or LNO approach to reduce the scaling with system size.

For the HLCs of non-Möbius structures, all methods can comfortably meet the 1 kcal mol$^{-1}$ threshold (see Table 7); DLPNO with tight settings can even reach down to 0.4 kcal mol$^{-1}$, and PNO-LCCSD(T) and LNO-CCSD(T) on their respective tight settings can go as low as 0.2 kcal mol$^{-1}$ RMSD. It is again the Möbius structures that are the most problematic for DLPNO-CCSD($T_x$) (x = 0,1) and PNO-LCCSD(T), while LNO-CCSD(T) is much more resilient to them. Even on default settings, LNO-CCSD(T) achieves RMSD = 0.7 kcal mol$^{-1}$ overall, which drops to 0.3 kcal mol$^{-1}$ on tight settings. PNO-LCCSD(T) on default settings and DLPNO-CCSD($T_1$) on TightPNO settings both come close to the 1 kcal mol$^{-1}$ mark overall. Particularly, the Möbius heptaphyrins throw a spanner in the works for DLPNO and PNO, which is much less the case for LNO on default settings, while no error exceeds 0.6 kcal mol$^{-1}$ for LNO on tight settings.

For the entire database of expanded porphyrins, we find an RMSD of 1.2–1.3 kcal mol$^{-1}$ both for PNO-LCCSD(T) on Normal settings and for DLPNO-CCSD($T_1$) on Tight settings.

The above results do make a good case for combining a localized HLC—for which either PNO-CCSD(T) Normal or DLPNO-CCSD($T_1$) Tight, but especially LNO-CCSD(T), would fit the bill—with a separate canonical MP2 calculation in a larger basis set—be it canonical RI-MP2 or DLPNO-MP2. For larger systems, eventually the $O(N^5)$ scaling of RI-MP2 would dominate the CPU time, but we have seen in Table 4 that especially DLPNO-MP2 with TightPNO can closely emulate canonical MP2 energetics. Another approach toward converging the MP2 part would be to carry out PNO-LMP2-F12 calculations.[63]

■ CONCLUSIONS

Localized natural orbital approaches are a very promising new alternative to both wave function methods and density functional theory. They, in principle, offer the gentle system size scaling of DFT without the empiricism (of accuracy) involved in the exchange-correlation functional—at the expense of introducing a measure of "empiricism of precision" through the various cutoffs introduced.

For systems with predominantly dynamical correlation, approaches like DLPNO-CCSD($T_1$) and PNO-LCCSD(T) seem to track canonical CCSD(T) results quite closely (see also the very recent paper[64] by Liakos, Guo, and Neese on the GMTKN55 benchmark suite[65]), while for truly severe static correlation, both canonical CCSD(T) and its localized approximations may be beyond help. Our results concern the intermediate regime, i.e., moderate static correlation: we found not only that discrepancies between canonical CCSD(T) and DLPNO-CCSD(T$_1$) or PNO-LCCSD(T) can reach several kcal mol$^{-1}$ for isomerization energies of chemical interest but that their magnitude is roughly proportional to several diagnostics for Type A static correlation. These problems can be somewhat mitigated by combining HLCs (i.e., CCSD(T)-MP2 differences), from the localized methods with more rigorous MP2 energetics, which are comparatively inexpensive to obtain. The LNO-CCSD(T) approach of Nagy and Kallay offers an alternative that, at least for the considered macrocycles, is more resilient to static correlation, especially with tight cutoffs, and can consistently approach the canonical values to better than 1 kcal mol$^{-1}$. It achieves this feat at the expense of CPU times being more dependent on the degree of static correlation. For the present problem, Möbius structures required about four times as much CPU time as their Hückel and figure-eight isomers, while the Möbius:Hückel ratio is only about 2:1 for the DLPNO and PNO approaches. As in so many scientific and nonscientific contexts, the TANSTAAFL principle[66] applies ("there ain't no such thing as a free lunch").

Table 7. [CCSD(T)-MP2] Relative Energies (kcal mol$^{-1}$) and Errors with Various Localized Orbital HLC Approximations for the Relative Energies of the Expanded Porphyrins under Consideration[a]

| HLC = [CCSD(T) − MP2] Basis: cc-pVDZ, p functions on H omitted | Canonical | DLPNO-CCSD(T$_0$) | | DLPNO-CCSD(T$_1$) | | PNO-LCCSD(T$_1$) | | LNO-CCSD(T) | | |
|---|---|---|---|---|---|---|---|---|---|---|
| | | Normal[b] | Tight[c] | Normal[b] | Tight[c] | Normal[d] | Tight[e] | Normal[f] | Tight[g] | Tight+[h] |
| 24H$_a$ | 0.6 | 0.9 | 0.3 | 0.9 | 0.2 | 0.8 | 0.3 | 0.2 | 0.1 | 0.1 |
| 24H$_b$ | 0.0 | 0.0 | 0.0 | 0.0 | 0.0 | 0.0 | 0.0 | 0.0 | 0.0 | 0.0 |
| 24M | 2.9 | 0.5 | 0.7 | 0.3 | 0.4 | 0.4 | 0.4 | -0.1 | 0.0 | 0.0 |
| 24TS$_1$ | -0.4 | 0.0 | 0.0 | 0.0 | 0.0 | 0.4 | -0.1 | 0.1 | 0.1 | 0.1 |
| 24TS$_2$ | 0.7 | 0.1 | 0.2 | 0.0 | 0.1 | 0.3 | 0.1 | -0.1 | 0.0 | 0.0 |
| 28F | 1.9 | -1.4 | -0.9 | -1.5 | -0.9 | -0.3 | -0.3 | -0.5 | -0.3 | -0.3 |
| 28M$_{1A}$ | 19.2 | 1.6 | 2.8 | 0.8 | 1.4 | 1.0 | 2.0 | -0.7 | -0.4 | -0.5 |
| 28M | 13.9 | 2.3 | 2.7 | 1.6 | 1.5 | 1.8 | 2.2 | -0.4 | -0.3 | -0.3 |
| 28M$_{1B}$ | 18.2 | 2.6 | 2.7 | 2.1 | 1.7 | 1.6 | 2.4 | -1.1 | -0.4 | -0.5 |
| 28TS$_3$ | -0.8 | -1.0 | -0.5 | -1.0 | -0.4 | 0.0 | 0.0 | -0.4 | -0.2 | -0.2 |
| 28R | 0.0 | 0.0 | 0.0 | 0.0 | 0.0 | 0.0 | 0.0 | 0.0 | 0.0 | 0.0 |
| 28TS$_{1A}$ | -2.1 | -0.3 | 0.0 | -0.2 | 0.2 | 0.3 | 0.2 | 0.0 | 0.1 | 0.1 |
| 28TS$_{1B}$ | -0.5 | -0.8 | -0.5 | -0.8 | -0.3 | -0.1 | -0.1 | -0.2 | -0.1 | -0.1 |
| 28TS$_{2A}$ | 20.2 | 2.5 | 3.4 | 1.2 | 1.9 | 1.6 | 2.4 | -0.5 | -0.2 | -0.3 |
| 28TS$_{2B}$ | 17.5 | 2.6 | 3.1 | 1.9 | 1.8 | 2.1 | 2.5 | -0.5 | -0.2 | -0.3 |
| 32F | 0.0 | 0.0 | 0.0 | 0.0 | 0.0 | 0.0 | 0.0 | 0.0 | 0.0 | 0.0 |
| 32H | -6.2 | 0.6 | 0.0 | 0.9 | 0.4 | -0.4 | -0.3 | 1.1 | 0.4 | 0.3 |
| 32TS$_2$ | -7.2 | 0.1 | -0.2 | 0.4 | 0.2 | -0.6 | -0.5 | 1.0 | 0.3 | 0.3 |
| 32M$_a$ | 6.6 | 4.5 | 3.3 | 4.0 | 2.5 | 2.1 | 2.4 | 1.3 | 0.6 | 0.2 |
| 32M$_b$ | 11.8 | 5.1 | 4.1 | 4.6 | 3.2 | 2.7 | 3.1 | 1.1 | 0.3 | 0.1 |
| 32TS$_1$ | -3.6 | 0.3 | -0.2 | 0.4 | 0.1 | -0.2 | -0.1 | 0.7 | 0.2 | 0.2 |
| RMSD | REF | 2.10 | 2.02 | 1.75 | 1.33 | 1.22 | 1.54 | 0.69 | 0.28 | 0.26 |
| Möbius | | 3.04 | 3.00 | 2.49 | 1.95 | 1.78 | 2.29 | 0.82 | 0.34 | 0.32 |
| Non-Möbius | | 0.72 | 0.39 | 0.74 | 0.38 | 0.40 | 0.24 | 0.57 | 0.23 | 0.19 |

[a]RMSDs from canonical results in the same basis set likewise in kcal mol$^{-1}$. Energy differences are heat-mapped on a continuous gradient from blue (most negative value) via white (zero) to red (most positive value). [b]NormalPNO (ORCA). [c]tightPNO (ORCA). [d]defaultDomain (MOLPRO). [e]tightDomain (MOLPRO). [f]lcorthr = normal (MRCC). [g]lcorthr = tight. [h]lcorthr = tight, wpairtol = 1 × 10$^{-6}$.

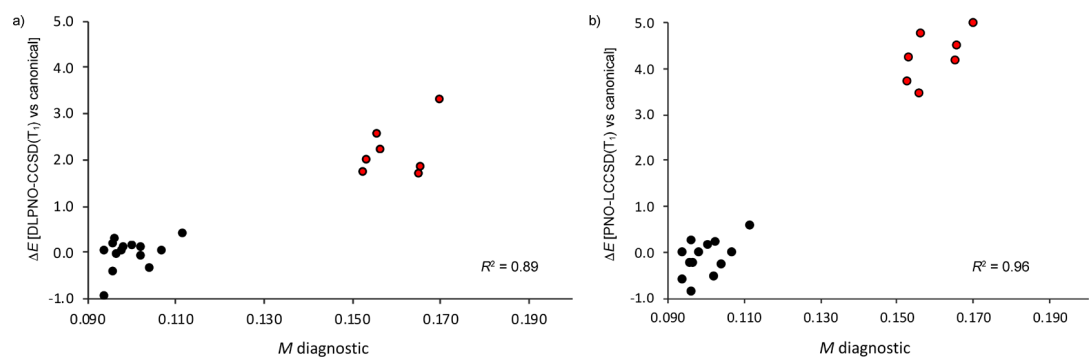

Figure 3. Relationship between the energy differences computed for various localized orbital CCSD(T) approximations and the canonical CCSD(T) method and the $M$ diagnostic for static correlation: a) For DLPNO-CCSD(T$_1$) with Tight settings and b) For PNO-LCCSD(T$_1$) with Tight settings. The Möbius structures are highlighted in red.

Finally, since the expanded porphyrins considered here and in ref 6 appear to be a useful test for resilience of quantum chemical approaches to static correlation, we propose the present POLYPYR21 data set as a benchmark for this purpose. The reference geometries, obtained at the B3LYP/6-311G-(d,p) level[67−69] in ref 6, are available for download as Supporting Information to the present paper.

## ASSOCIATED CONTENT

### Supporting Information

The Supporting Information is available free of charge at https://pubs.acs.org/doi/10.1021/acs.jctc.0c00297.

Geometries for all structures included in the POLY-PYR21 data set provided in .xyz format, and in addition, total and relative energies at all levels considered are provided in a Microsoft Excel spreadsheet (ZIP)

## AUTHOR INFORMATION

### Corresponding Authors

Jan M. L. Martin − *Department of Organic Chemistry, Weizmann Institute of Science, 76100 Reḥovot, Israel;* orcid.org/0000-0002-0005-5074; Email: gershom@weizmann.ac.il; Fax: +972 8 9343029

Mercedes Alonso − *Department of General Chemistry (ALGC), Vrije Universiteit Brussel (VUB), 1050 Brussels, Belgium;* orcid.org/0000-0002-7076-2305; Email: mercedes.alonso.giner@vub.be

### Authors

Nitai Sylvetsky − *Department of Organic Chemistry, Weizmann Institute of Science, 76100 Reḥovot, Israel*

Ambar Banerjee − *Department of Organic Chemistry, Weizmann Institute of Science, 76100 Reḥovot, Israel;* orcid.org/0000-0001-6113-7033

Complete contact information is available at:
https://pubs.acs.org/10.1021/acs.jctc.0c00297

### Author Contributions

[†](N.S. and A.B.) Equally contributing first authors.

### Funding

M.A. thanks the FWO for a postdoctoral fellowship (12F4416N) and the VUB for financial support. Research at Weizmann was funded by the Israel Science Foundation (Grant 1358/15) and by the Estate of Emile Mimran (Weizmann), while computational resources and services were provided by Chemfarm (the Weizmann Institute Faculty of Chemistry HPC facility). A.B. and N.S. acknowledge postdoctoral and doctoral fellowships, respectively, from the Feinberg Graduate School at the Weizmann Institute.

### Notes

The authors declare no competing financial interest.

## ACKNOWLEDGMENTS

The authors would like to thank Prof. Hans-Joachim Werner (U. of Stuttgart, Germany) for helpful discussions and Dr. Peter Nagy (TU Budapest, Hungary) for extensive assistance with the MRCC program system. Dr. Mark Vilensky (scientific computing manager of ChemFarm) is thanked for his assistance with the somewhat exorbitant mass storage requirements of the largest canonical calculations.

## ■ NOTE ADDED IN PROOF

According to LNO-CCSD(T) vtight calculations on 32F, 32Ma, and 32Mb (P. Nagy, personal communication), both 32Ma and 32Mb are 17.6 kcal/mol above 32H, i.e., at 0.8 and 0.9 kcal/mol, respectively, above the canonical values, nearly identical to the more economical Tight+ results of 0.8 and 0.7 kcal/mol, respectively.